\documentclass[fleqn,usenatbib]{mnras}

\usepackage{mathptmx}

\usepackage[T1]{fontenc}

\usepackage{graphicx}	
\usepackage{xspace}
\usepackage{amsmath}	
\usepackage{threeparttable}
\usepackage{array}
\usepackage{xcolor}
\usepackage{multirow}
\usepackage{bold-extra}

\graphicspath{{./}{figures/}}

\title[Baikal-GVD neutrino from TXS~0506+056]{
High-energy neutrino-induced cascade from the direction of the flaring radio blazar TXS~0506+056 observed by Baikal-GVD in 2021}

\author[Baikal-GVD collaboration et al.]{\parbox{\textwidth}{
\large
V.~A.~Allakhverdyan,$^{1}$
A.~D.~Avrorin,$^{2}$
A.~V.~Avrorin,$^{2}$
V.~M.~ Aynutdinov,$^{2}$
Z.~Barda\v{c}ov\'{a},$^{3,4}$
I.~A.~Belolaptikov,$^{1}$
E.~A.~Bondarev,$^{2}$
I.~V.~Borina,$^{1}$
N.~M.~Budnev,$^{5}$
V.~A.~Chadymov,$^{6}$
A.~S.~Chepurnov,$^{7}$
V.~Y.~Dik,$^{1,8}$
G.~V.~Domogatsky,$^{2}$
A.~A.~Doroshenko,$^{2}$
R.~Dvornick\'{y},$^{3}$
A.~N.~Dyachok,$^{5}$
Zh.-A.~M.~Dzhilkibaev,$^{2}$
E.~Eckerov\'{a},$^{3,4}$
T.~V.~Elzhov,$^{1}$
L.~Fajt,$^{4}$
V.~N.~Fomin,$^{6}$
A.~R.~Gafarov,$^{5}$
K.~V.~Golubkov,$^{2}$
N.~S.~Gorshkov,$^{1}$
T.~I.~Gress,$^{5}$
K.~G.~Kebkal,$^{9}$
I.~Kharuk,$^{2}$
E.~V.~Khramov,$^{1}$
M.~M.~Kolbin,$^{1}$
S.~O.~Koligaev,$^{10}$
K.~V.~Konischev,$^{1}$
A.~V.~Korobchenko,$^{1}$
A.~P.~Koshechkin,$^{2}$
V.~A.~Kozhin,$^{7}$
M.~V.~Kruglov,$^{1}$
V.~F.~Kulepov,$^{11}$
Y.~E.~Lemeshev,$^{5}$
M.~B.~Milenin,$^{2}$\thanks{Deceased}
R.~R.~Mirgazov,$^{5}$
D.~V.~Naumov,$^{1}$
A.~S.~Nikolaev,$^{7}$
D.~P.~Petukhov,$^{2}$
E.~N.~Pliskovsky,$^{1}$
M.~I.~Rozanov,$^{12}$
E.~V.~Ryabov,$^{5}$
G.~B.~Safronov,$^{2}$
D.~Seitova,$^{1,8}$
B.~A.~Shaybonov,$^{1}$
M.~D.~Shelepov,$^{2}$
S.~D.~Shilkin,$^{2}$
E.~V.~Shirokov,$^{7}$
F.~\v{S}imkovic,$^{3,4}$
A.~E. Sirenko,$^{1}$
A.~V.~Skurikhin,$^{7}$
A.~G.~Solovjev,$^{1}$
M.~N.~Sorokovikov,$^{1}$
I.~\v{S}tekl,$^{4}$
A.~P.~Stromakov,$^{2}$
O.~V.~Suvorova,$^{2}$
V.~A.~Tabolenko,$^{5}$
B.~B.~Ulzutuev,$^{1}$
Y.~V.~Yablokova,$^{1}$
D.~N.~Zaborov,$^{2}$
S.~I.~Zavyalov,$^{1}$
D.~Y.~Zvezdov$^{1}$
(Baikal-GVD Collaboration), \\
A.~K.~Erkenov,$^{13}$
N.~A.~Kosogorov,$^{14,15,16}$
Yu.~A.~Kovalev,$^{14}$
Y.~Y.~Kovalev,$^{17,14,15}$\thanks{Corresponding author, e-mail: yykovalev@gmail.com}
A.~V.~Plavin,$^{18,14}$
A.~V.~Popkov,$^{15,14}$
A.~B.~Pushkarev,$^{19,14}$
D.~V.~Semikoz,$^{20}$
Y.~V.~Sotnikova,$^{13}$
S.~V.~Troitsky$^{2,21}$}
\vspace{0.3cm}\\
\parbox{\textwidth}{
$^{1}$Joint Institute for Nuclear Research, Dubna, 141980  Russia\\
$^{2}$Institute for Nuclear Research of the Russian Academy of Sciences, 60th October Anniversary Prospect 7a, Moscow 117312, Russia\\
$^{3}$Comenius University, Bratislava, 81499 Slovakia\\ 
$^{4}$Czech Technical University in Prague, Institute of Experimental and Applied Physics, 11000 Prague, Czech Republic\\
$^{5}$Irkutsk State University, Irkutsk, 664003 Russia\\ 
$^{6}$Independent researcher\\
$^{7}$Skobeltsyn Research Institute of Nuclear Physics, Lomonosov Moscow State University, Moscow, 119991 Russia\\
$^{8}$Institute of Nuclear Physics ME RK, Almaty, 050032 Kazakhstan\\
$^{9}$LATENA, St. Petersburg, 199106, Russia\\ 
$^{10}$INFRAD, Dubna, 141981, Russia\\
$^{11}$Nizhny Novgorod State Technical University, Nizhny Novgorod, 603950 Russia\\ 
$^{12}$Sankt Petersburg State Marine Technical University, Sankt Petersburg, 190008 Russia\\ 
$^{13}$Special Astrophysical Observatory of RAS, Nizhny Arkhyz 369167, Russia\\
$^{14}$Astro Space Center of Lebedev Physical Institute, Profsoyuznaya 84/32, 117997 Moscow, Russia\\
$^{15}$Moscow Institute of Physics and Technology, Institutsky per. 9, Dolgoprudny 141700, Russia\\
$^{16}$Cahill Center for Astronomy and Astrophysics, MC 249-17 California Institute of Technology, Pasadena, CA 91125, USA\\
$^{17}$Max-Planck-Institut f\"ur Radioastronomie, Auf dem H\"ugel 69, 53121 Bonn, Germany\\
$^{18}$Black Hole Initiative at Harvard University, 20 Garden Street, Cambridge, MA 02138, USA\\
$^{19}$Crimean Astrophysical Observatory, Nauchny 298409, Crimea, Russia\\
$^{20}$APC, Universit\'e Paris Diderot, CNRS/IN2P3, CEA/IRFU, Sorbonne Paris Cit\'e, 119 75205 Paris, France\\
$^{21}$Physics Department, M.V. Lomonosov Moscow State University, 1-2 Leninskie Gory,  Moscow 119991, Russia
}}

\date{Accepted 2023 November 21. Received 2023 November 13; in original form 2023 September 20}

\pubyear{2023}

\begin{document}
\label{firstpage}
\pagerange{\pageref{firstpage}--\pageref{lastpage}}
\maketitle

\begin{abstract}
\normalsize
The existence of high-energy astrophysical neutrinos has been unambiguously demonstrated, but their sources remain elusive. IceCube reported an association of a 290-TeV neutrino with a gamma-ray flare of TXS~0506+056, an active galactic nucleus with a compact radio jet pointing to us. Later, radio-bright blazars were shown to be associated with IceCube neutrino events with high statistical significance. These associations remained unconfirmed with the data of independent experiments. Here we report on the detection of a rare neutrino event with the estimated energy of $224\pm75$~TeV from the direction of TXS~0506+056 by the new 
Baikal Gigaton Volume Detector (Baikal-GVD) in April 2021. This event is the highest-energy cascade detected so far by the Baikal-GVD neutrino telescope from a direction below horizon. The result supports previous suggestions that radio blazars in general, and TXS~0506+056 in particular, are the sources of high-energy neutrinos, and opens up the cascade channel for the neutrino astronomy.
\end{abstract}


\section{Introduction}\label{s:intro}

Recently, the Baikal-GVD neutrino telescope collaboration reported \citep{Baikal-JETP, Baikal-PRD2023, 2023MNRAS.526..942A} on its first detection of astrophysical neutrinos with hundreds of TeV energies. This gives the first independent confirmation of the results of the IceCube telescope that has been observing high-energy neutrinos since 2008 \citep{IceCubeFirst26}. Yet, astrophysical neutrino origins, as well as the details of their production mechanisms, still remain poorly understood, see e.g.~\citet{ST-UFN} for a review.

Active galactic nuclei (AGNs) have been considered as potential neutrino sources since the very early days of multi-messenger astronomy. Recent statistical observational studies associate high-energy neutrino emission with blazars, a class of active galaxies with jets pointing towards the observer \citep{neutradio1,neutradio2,neutradio3,2020MNRAS.497..865G,Hovatta2021,2022ApJ...933L..43B,2023arXiv230511263B,Alisa2024,ANTARES_RFC}.
Note that different authors employ different methods and use different neutrino data sets to analyze for possible blazar-neutrino connection, arriving to strongly varying significance estimates. While the original claims of the association are  confirmed when the same data are used, see e.g.\ \citep{Anna2022,IC-VLBI2023,2023ApJ...955L..32B}, newer reconstructions of IceCube data or different selection procedures may result in the apparent reduction of the association significance \citep{2021PhRvD.103l3018Z,IC-VLBI2023,2023ApJ...955L..32B}. This trend was reported as well for the original TXS~0506$+$056 association by \citet{IceCube-TXS0506-2023}, for various blazar statistical associations by \citet{2023ApJ...955L..32B}, and for potential connection of high-energy neutrinos with tidal disruption events by \citet{IceCube-TDE-ICRC2023}; the reasons for this remain to be understood. 

Like photons, neutrinos experience relativistic beaming in blazars, facilitating their detection on Earth. Furthermore, evidence was found for neutrinos being preferentially produced during major flares in the central parsecs of blazars \citep{IceCubeTXSgamma,neutradio1,Hovatta2021}. What are the specific regions where neutrinos are produced and what is the physical connection between neutrinos and electromagnetic flares, remain open questions \citep[see e.g.][for discussion of models]{Neronov-model,Polina-cores}.

We note that neutrino sky is complex and diverse.
Recently, IceCube has found neutrino emission from the nearby galaxy which does not posses a relativistic jet, NGC~1068 \citep{NGC_IC}.
A significant component of high energy neutrino emission is now also discovered from our Galaxy \citep{neutgalaxy,ANTARES-ridge,IC_Galaxy2023}, see also discussion in \citet{2023MNRAS.526..942A}.
As a result, the fraction of neutrino flux to arrive from blazars is still strongly debated.

The first association of an individual astrophysical object with a high-energy neutrino was found for the TXS~0506+056 blazar \citep{IceCubeTXSgamma}. On September 22, 2017, the IceCube observatory detected a probable 290-TeV neutrino from the direction of this blazar. TXS~0506+056 was experiencing a major flare across the entire electromagnetic spectrum, from radio to gamma rays.
Follow up observations of TXS~0506+056 were performed by the ANTARES neutrino telescope \citep{ANTARES:2018osx}. ANTARES has detected one low-energy muon neutrino event from the direction of the blazar which was consistent with the background.

Since then, this object has been extensively studied using astronomical instruments operating at various electromagnetic wavelengths. 
This object is characterized by bright compact emission on parsec scales, but appears to be completely ordinary otherwise \citep{2020AdSpR..65..745K}. Further studies of neutrino emission of TXS~0506+056 can shed more light on the mechanisms of neutrino production, not only in this particular source but in the whole bright blazar population.

\section{
Baikal-GVD detection of a high energy neutrino from the direction of TXS 0506+056
} 
\label{s:obsBaikal}

Observation of astrophysical high-energy neutrinos
requires detectors of very large volume because neutrino interacts with matter very weakly. 
The Baikal Gigaton Volume Detector, or Baikal-GVD, 
is a Northern-hemisphere cubic-kilometer scale deep underwater Cherenkov detector aimed at the search for neutrinos with energies\footnote{1~TeV$=10^{12}$~eV; 1~PeV$=10^{15}$~eV.} between several TeV and tens of PeV. The telescope is located in the Southern part of Lake Baikal ($51^\circ50^\prime$ N, $104^\circ20^\prime$ E), at about 4 km from the shore. The lake depth at the facility site is 1366 m. The instrument has been operating since 2016 with the operational volume increasing every year in the course of the detector deployment.

Neutrinos are detected through the Cherenkov radiation emitted by secondary particles produced in neutrino interactions in the water or bedrock below the detector and observed by the detector light sensors, optical modules (OMs). Charged-current (CC) muon neutrino interactions yield long-lived muons that can pass several kilometers through the water, leading to a track signature crossing the detector, see e.g.~\citet{Baikal-GVD:tracks}. The accuracy of the angular reconstruction better than $1^\circ$ is typical of high energy track-like events in neutrino telescopes. Tracks are therefore often used for the search of point sources of astrophysical neutrinos. However, the precision of the determination of the neutrino energy is poor in this case.

\begin{table*}
	\centering
	\caption{Reconstructed parameters of the highest-energy upgoing cascade event: arrival time (Modified Julian Day, MJD), reconstructed energy $E_{\rm sh}$, number of hits $N_{\rm hit}$, zenith angle $\theta$, right ascension (RA) and declination (Dec), Galactic longitude $l$ and latitude $b$.
}
	\label{tab:event}
	\begin{tabular}{ccccccccc}
	\hline
	Event ID     & MJD    & $E_{\rm sh}$, & $N_{\rm hit}$& $\theta$, & RA,   & Dec,  & l,   & b,   \\
	               &  &          (TeV) &     & (deg)   & (deg) & (deg) & (deg) & (deg) \\
	\hline
	GVD210418CA     & 59322.94855324 & 224$\pm$75 &  24  & 115     & 82.4  & 7.1   & 163.2   & $-14.6$    \\
	\hline
	\end{tabular}
\end{table*}

Neutral-current (NC) interactions, as well as most of CC interactions of electron and tau neutrinos, yield hadronic and electromagnetic showers (cascades). The showers are quasi point-like, highly anisotropic
sources of Cherenkov radiation.
The energy of the cascade progenitor neutrino is determined with a good accuracy;
however, the angular reconstruction is worse for cascades than for tracks. The cascade channel is thus complementary to the track one. Previously, cascades were used mostly in searches for the diffuse neutrino flux and in studies of the astrophysical neutrino spectrum.

Optical properties of Baikal deep water are characterized by the light absorption length $L_\mathrm{a} \approx (21-23)$~m and the scattering length $L_\mathrm{s} \approx (60-80)$~m at $\lambda = (480-500)$~nm wavelength. Compared to the ice used as the target medium in the IceCube neutrino telescope, this low-scattering medium makes it possible to reconstruct the arrival directions of cascade progenitor neutrinos in Baikal-GVD with a much better accuracy, which varies from one event to another but is of order a few degrees, compared to tens of degrees in ice. This opens the previously unavailable possibility to use the cascade channel for searches for neutrino point sources.

\begin{figure}
    \centering
    \includegraphics[width=0.9\linewidth]{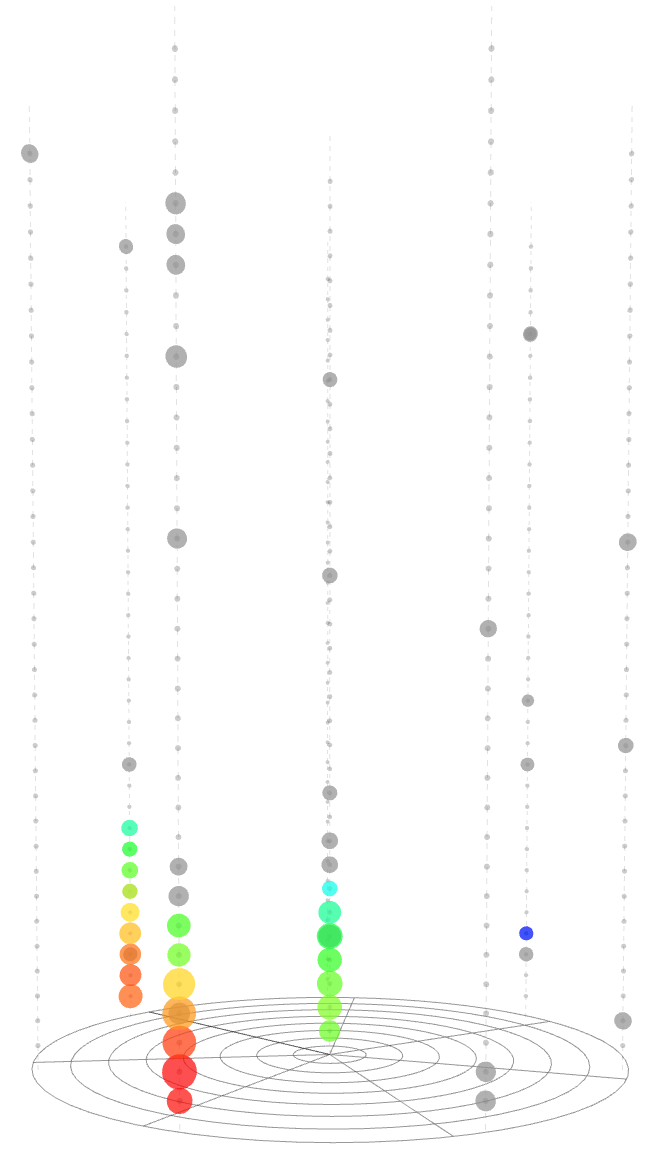}
    \caption{Image of the response of the Baikal-GVD detector to the GVD210418CA event.
    The color gradient corresponds to the earlier (red) and later (blue) photon arrival time of hit optical modules, while the size of the spheres is a measure for the recorded number of photo-electrons. Grey points represent hits excluded from the analysis (noise hits).}
    \label{f:fig1}
\end{figure}

A distinctive feature of the Baikal-GVD design is its modular structure, which allows performing physical studies even at early stages of setup deployment. The telescope is composed of several sub-arrays, clusters of optical modules.
Each cluster is an independent array comprising a total of 288 OMs and is connected to the shore station by its own electro-optical cable. 
The current rate of array deployment is about two clusters per year, see for details \autoref{a:Baikal}.

In this work, we use the approach to search for high-energy neutrinos of astrophysical origin based on the selection of events from showers generated in the sensitive volume of the Baikal-GVD neutrino telescope. The procedures of shower-like event selection and cascade reconstruction by a cluster of Baikal-GVD are presented elsewhere \citep[e.g.][]{Baikal-JETP,Baikal-ICRCPOS2021}. 
The reconstruction precision of the cascade vertex position is about 0.5\,m\,--\,1\,m. It affects very weakly the precision of the cascade direction and energy reconstruction.
The precision of the energy reconstruction is about 10\,\%\,--\,30\,\% and depends substantially on the energy of the cascade, its position and orientation relative to the cluster. The precision of reconstruction of the progenitor neutrino direction depends strongly on OM hit multiplicity and is about $2^\circ$\,--\,$4^\circ$ (median value). In this analysis, we use the Baikal-GVD data collected between April, 2018, and March, 2022, which corresponds to 13.5 years of equivalent live time for one cluster. The sample of 3.49$\times$10$^{10}$ events was collected by the basic trigger of the telescope. After  noise-hit suppression, cascade reconstruction and application of cuts on the reconstruction quality parameters, a sample of 14328 cascades with reconstructed energy $E_\mathrm{sh}>10$~TeV and OM hit multiplicity $N_\mathrm{hit}>$11 was selected. Most of these events have atmospheric origin and constitute the background for the astrophysical neutrino search.

Additional, drastic background suppression is achieved by selecting only upward moving cascades, as it was 
described in \citet{Baikal-PRD2023}. Cascade-like events with $E_\mathrm{sh}>15$~TeV and reconstructed zenith angle $\theta$ satisfying $\cos\theta < -0.25$, were selected as astrophysical neutrino candidates. A total of 11 such events have been found in the data sample, while on average, 2.7 events are expected from atmospheric neutrinos and 0.5 events from mis-reconstructed atmospheric muons. 

\begin{figure}
    \centering
    \includegraphics[width=0.9\linewidth]{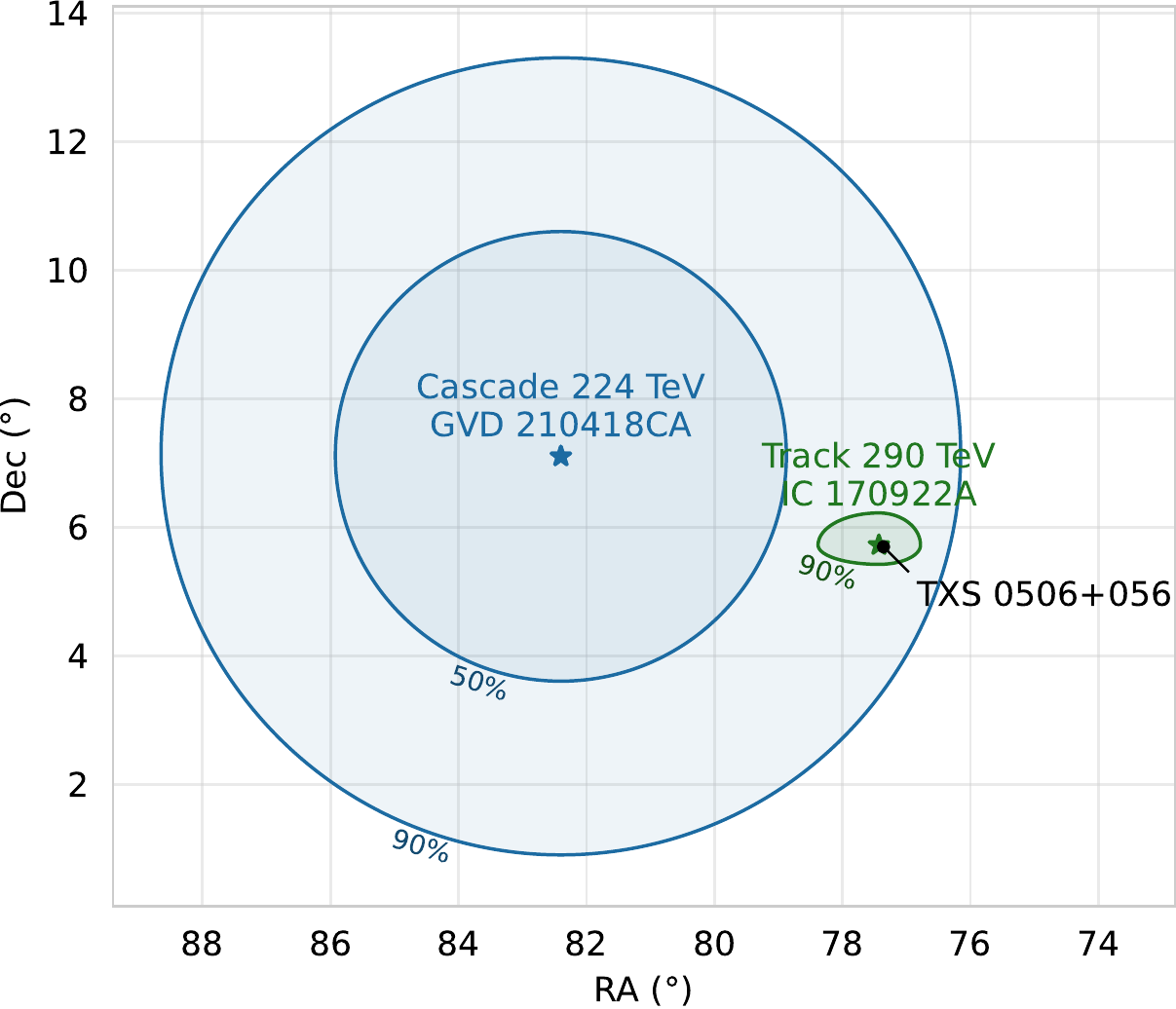}
    \caption{Arrival directions of the IceCube 2017 track event and the Baikal-GVD 2021 cascade event with their directional uncertainties, together with the position of TXS~0506+056.}
    \label{f:fig2}
\end{figure}

The highest-energy cascade detected so far by Baikal-GVD from a direction below the horizon arrived on April 18, 2021. The basic parameters of this event are presented in \autoref{tab:event}.
The pattern of this event in the Baikal-GVD telescope is show in \autoref{f:fig1}.

For the given $E_\mathrm{sh}$,  $N_\mathrm{hit}$, and shower direction, the probability that this event comes from the atmospheric background can be estimated from Monte-Carlo simulations to be as low as $P_\mathrm{atm} = 0.0033$.
This probability refer to the number of expected background events with $E_\mathrm{sh} \geq $ 224 TeV, $N_\mathrm{hit} \geq $ 24 and zenith angle $\theta \geq$ $115^\circ$.
Assuming $E^{-2.46}$ spectrum of astrophysical neutrinos \citep{IceCube:HESE2020}, we determine the ``signalness'' parameter \citep{IceCubeOldAlerts}, which is a measure of the likelihood of the event to have the astrophysical origin relative to the background. The signalness of the GVD210418CA cascade is 97.1\%. For comparison, the signalness of the track event IceCube-170922A, associated with TXS~0506+056, is only 56.5\% \citep{IceCubeTXSgamma}. The Monte-Carlo simulations of this event demonstrate that in 90\% of cases, the reconstructed neutrino arrival direction is within the opening angle of $6.2^\circ$ from the assumed one. This 90\%-containment circle contains the direction to the TXS~0506+056 blazar, as illustrated in \autoref{f:fig2}. Below we discuss the possibility that the highest-energy Baikal-GVD upgoing cascade, GVD210418CA, is indeed associated with this well-known blazar. 

\section{Details of neutrino and electromagnetic observations and data processing}

\subsection{Baikal-GVD observations of cascade events} \label{a:Baikal}

\begin{figure}
    \centering
    \includegraphics[width=\linewidth,trim=1.2cm 1.5cm 0.6cm 1cm]{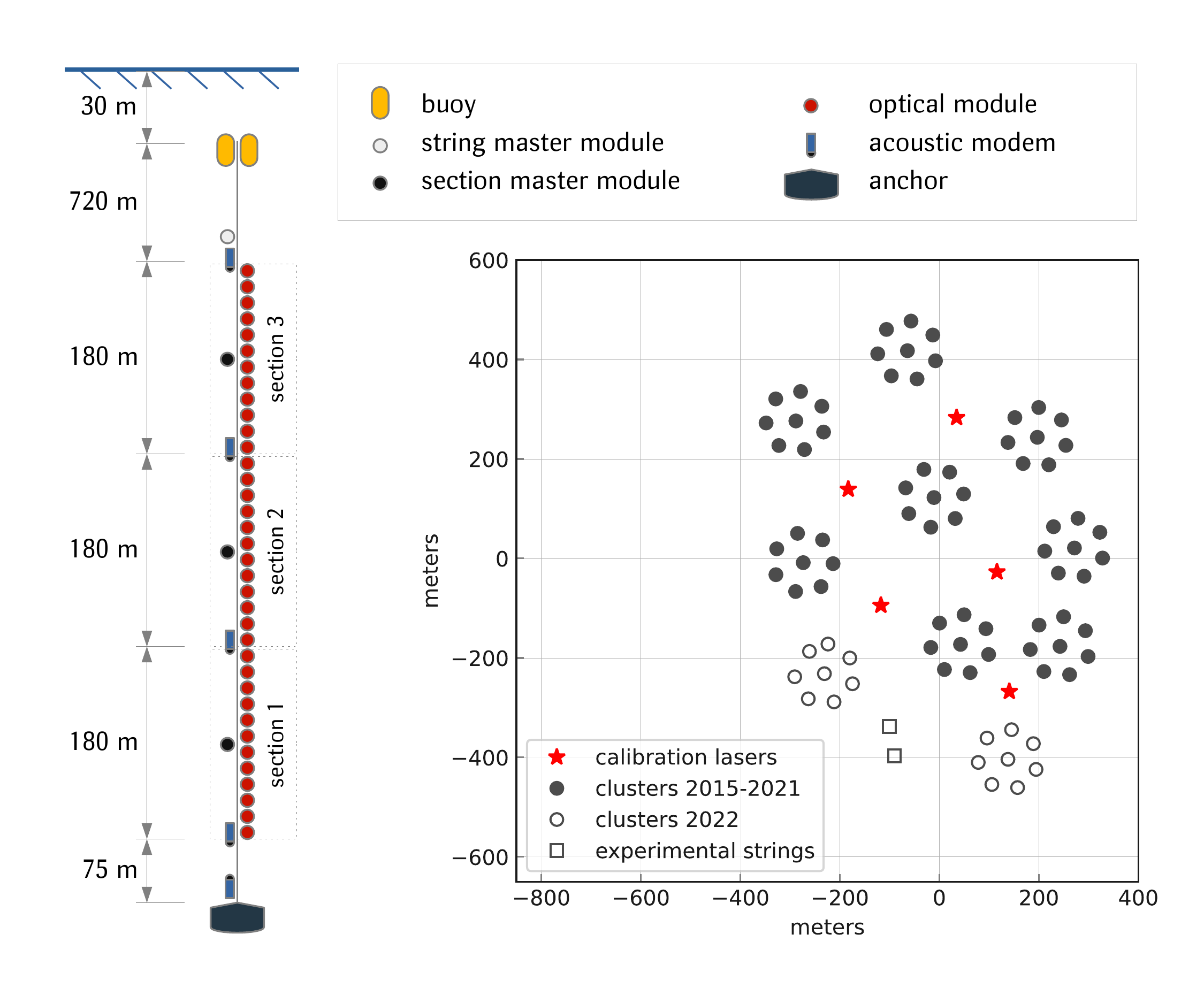}
    \caption{Scheme of Baikal-GVD. 
    \textit{Left:} string structure (side view). 
    \textit{Right:} cluster arrangement scheme (top view).}
    \label{f:SA}
\end{figure}

The Baikal-GVD neutrino telescope has a modular configuration with a basic structure named cluster. Each cluster is operated as an independent detector, taking data upon its deployment \citep{Baikal-YADFIZ:2022}.
A standard cluster with eight strings (one central and seven peripheral strings) of 288 OMs in total covers a cylindrical instrumental volume of 525~m in height and 120~m in diameter. A basic element of the cluster is an OM equipped with a 25~cm photo-multiplier tube and related electronics. Inter-cluster distances measured between the central strings vary between 250~m and 300~m. The choice of the geometrical  parameters follows the results of a Monte Carlo study optimizing the high-energy neutrino detection rate at trigger level. It implies a balance between two factors. On the one hand, large distances between photo detectors are needed to obtain a larger effective volume. On the other hand, sufficient density of OMs is required to accurately reconstruct the neutrino interaction vertex, energy, and direction. The operational configurations of Baikal-GVD in 2021 (8 clusters, 64 strings, 2304 OMs) and 2022 (10 clusters, 80 strings, 2880 OMs) are shown schematically in \autoref{f:SA}. 

Luminescence of organic particles suspended in water is a natural background for the detection of Cherenkov photons from charged particles by OMs.
The luminescence results in typical OM illumination at the level of one photo-electron (ph.el.). The cut on the lowest allowed charge of OM hits $Q > 1.5$~ph.el.\ allows suppressing the number of noise pulses from water luminescence substantially.
At the time of recording the event from the direction of TXS~0506+056, the level of noise rate of photons from water luminescence was low, as usually happens in April every year. 
The monitoring of the water optical properties in April 2021 indicates an absorption length of about 22~m --- very close to the long-term average --- and a relatively high value of the scattering length, more than 60 m. Thus the event we discuss in this paper was observed under favorable operational conditions.

\begin{figure}
    \centering
    \includegraphics[trim=1cm 0cm 2.5cm 0cm,width=\linewidth]{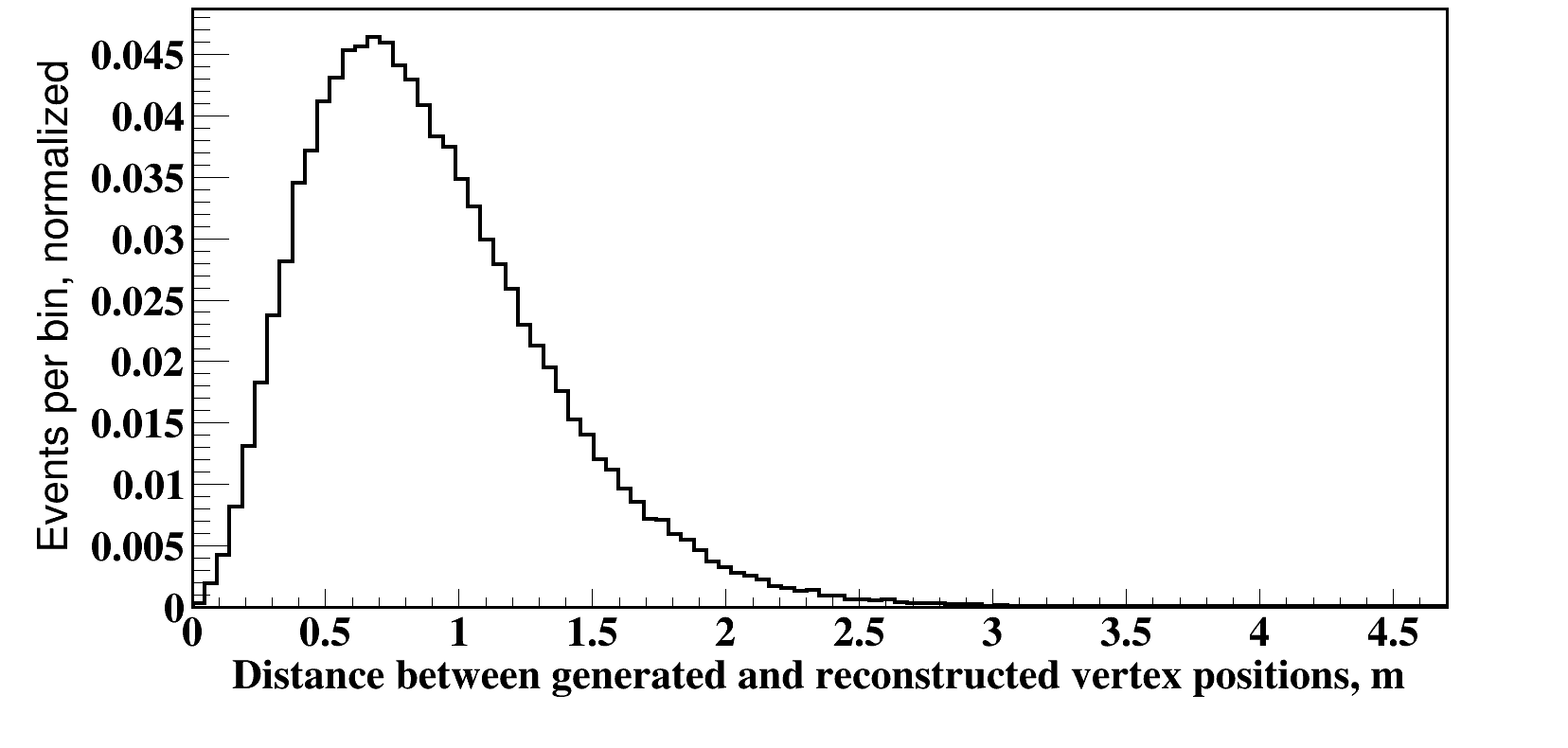}
    \includegraphics[trim=1cm 0cm 3cm 0cm,width=\linewidth]{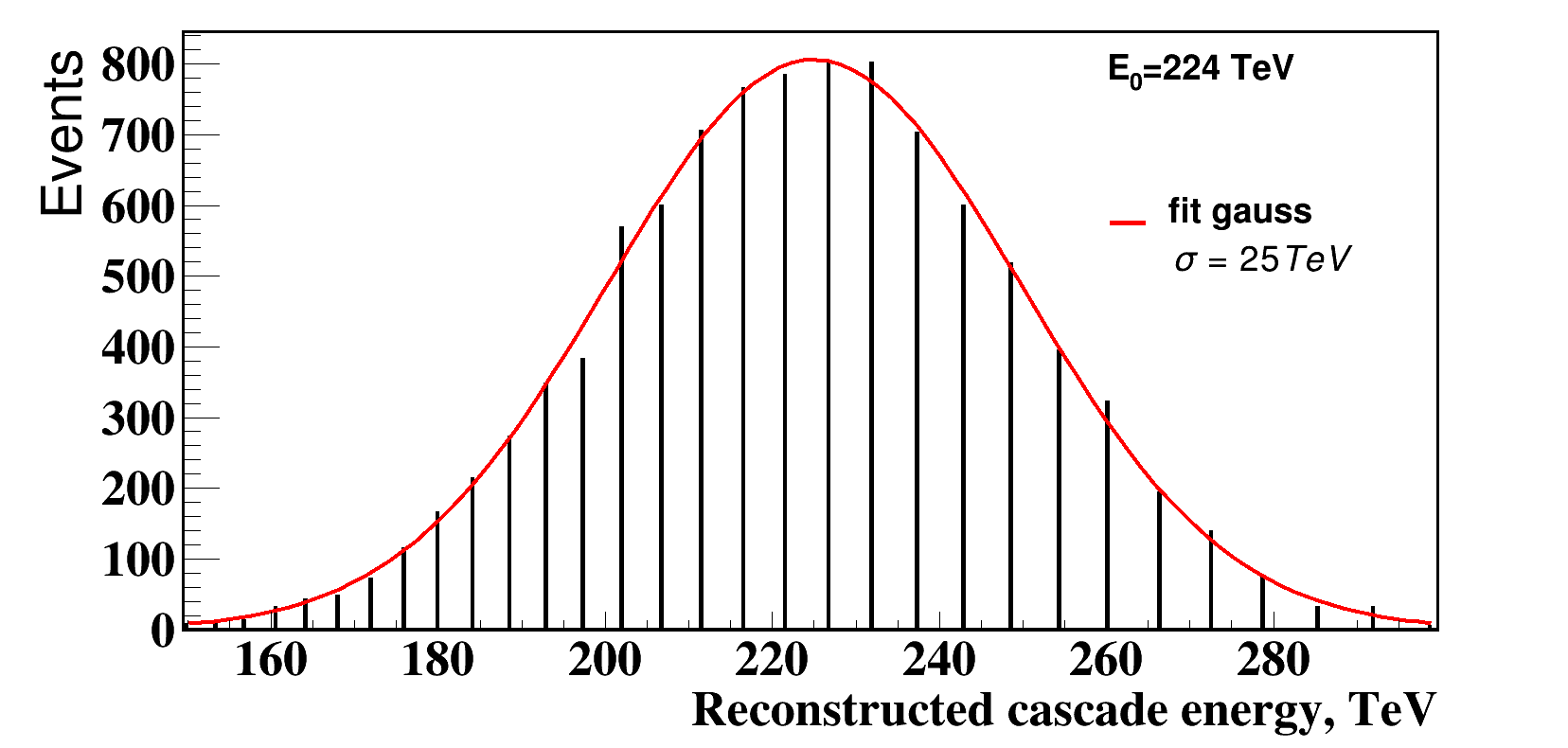}
    \includegraphics[trim=0cm 0cm -0.7cm 0cm,width=\linewidth]{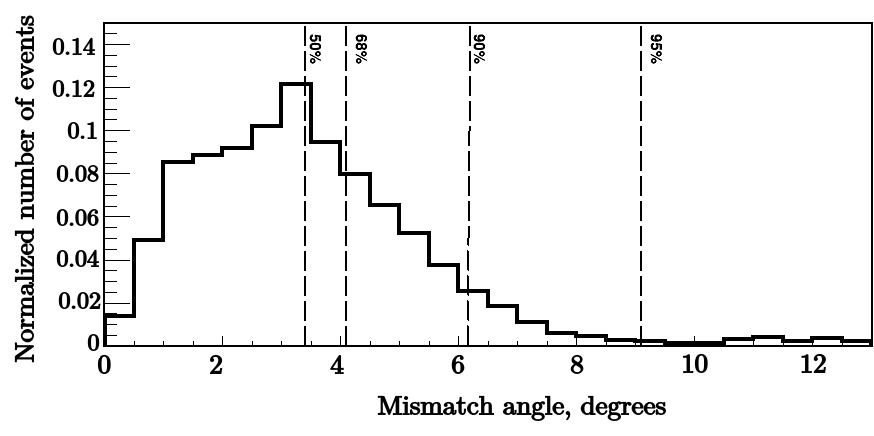}
    \caption{
    \textit{Top:} Distribution of the distance between a fixed vertex position of the generated cascade and the reconstructed vertex position.
    \textit{Middle:} Distribution of the reconstructed energy of Monte-Carlo cascades with a fixed energy of the generated cascade 224~TeV; the fitted Gaussian $\sigma=25$~TeV.
    \textit{Bottom:} Distribution of Monte-Carlo simulated differences between generated and reconstructed arrival directions. The 50\,\% mismatch angle is $3.4^\circ$, for 90\,\% we find $6.2^\circ$.
    This was obtained taking into account the uncertainty in the recovered energy of the event.
    \label{f:SB}}
\end{figure}

The Cherenkov radiation from electromagnetic and hadronic showers is formed by photons emitted by charged particles of the shower (mainly by electrons and positrons) and is determined by their spatial, angular, and temporal distributions. Cascades are reconstructed as point-like sources of light. The coordinates, energy, and orientation of showers are reconstructed in two stages. At the first stage, the shower coordinates are reconstructed using temporal information from the telescope’s triggered OMs (in the approximation of a point source). The reconstruction procedure consists in determining of the shower coordinates that correspond to the minimum of the $\chi^2$ function. At the second stage, the shower energy and direction are reconstructed by the maximum likelihood method using the shower coordinates reconstructed at the first stage \citep{Baikal-AstronL:2009}. 
The precision of the reconstruction of the arrival direction of the neutrino, which caused the GVD210418CA cascade, was estimated by the reconstruction of Monte-Carlo generated cascades with
vertices positions, energies and directions equal to reconstructed values of the event. This takes into account the uncertainty in the vertex position and the energy. 
The distribution of the distance between a fixed vertex position of the generated cascade and the reconstructed vertex position, as well as reconstructed cascade energy and the mismatch angle between generated and reconstructed directions are shown in \autoref{f:SB}.

\subsection{RATAN-600 multi-frequency radio monitoring} \label{a:RATAN}

The 600-meter ring radio telescope RATAN-600 of the Special Astrophysical Observatory \citep{KP79}, the Russian Academy of Sciences, performs long-term observations of continuum spectra at centimeter wavelengths of a large sample of extragalactic radio sources selected with very long baseline interferometry (VLBI) technique \citep[e.g.][]{1999A&AS..139..545K,2002PASA...19...83K}.

\begin{figure}
   \centering
   \includegraphics[width=0.95\linewidth]{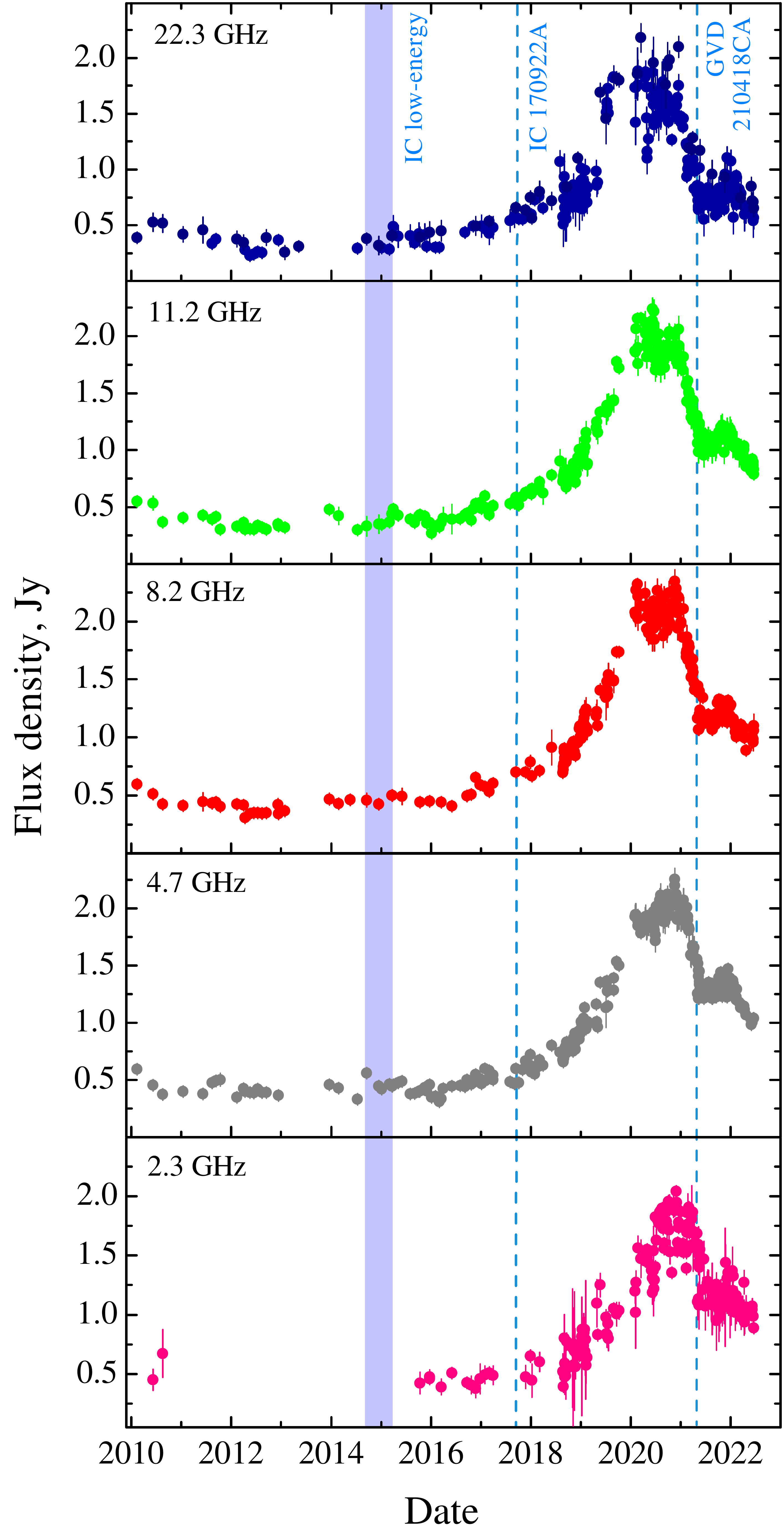}
   \caption{RATAN-600 light curves of TXS 0506+056 measured at 2.3-22.3 GHz in 2010.0--2022.5. The blue vertical lines mark different IceCube and Baikal-GVD neutrino events which are labeled in the plot.}
  \label{f:RATANall}
\end{figure}

The RATAN-600 measurements of TXS~0506+056 
were carried out at frequencies of 2.3, 4.7, 8.2, 11.2, and 22.3~GHz in 2010-2022 (\autoref{f:RATANall}). The earlier RATAN-600 instantaneous continuum spectra since 1997 are published in \citet{2020AdSpR..65..745K}. The observations were made in the RATAN transit mode, and the multi-frequency broad-band radio spectra were obtained at the five above-mentioned frequencies within 3-5 min. The radio telescope parameters are described in \citet{1993IAPM...35....7P,
1999A&AS..139..545K,
2000PASJ...52.1027K,2020gbar.conf...32S}.
The measurements were processed using the Flexible Astronomical Data Processing System (FADPS) standard package modules \citep{1997ASPC..125...46V} for the RATAN continuum radiometers and the automated data reduction systems \citep{2011AstBu..66..109T,2018AstBu..73..494T,2016AstBu..71..496U,1999A&AS..139..545K}. We used the following flux density secondary calibrators: 3C\,48, 3C\,138, 3C\,147, 3C\,161, 3C\,286, and NGC~7027. 
The calibrator measurements were corrected for the angular size and linear polarization, according to the data from \citet{1994A&A...284..331O} and \citet{1980A&AS...39..379T}.
The flux densities of the amplitude calibrators were used within the scale by \citet{1977A&A....61...99B}, their temporal evolution was considered following \citet{1994A&A...284..331O} and \citet{2013ApJS..204...19P,2017ApJS..230....7P}.

\subsection{Flare decomposition of radio light curves} \label{a:flares}

It is often assumed that radio flux density variations in blazars are due to the features forming and flaring in the jet and a contribution from the underlying jet \citep[e.g.][]{1999ApJS..120...95V}.
We decomposed our radio light curves into a few exponential flares and a constant quiescent level, $\Delta S_0$.  A growth and decay of each flare was modeled in the following form \citep[e.g.][]{2009A&A...494..527H}:
$$
    \Delta S(t)=\left\{
                \begin{array}{ll}
                  \Delta S_\mathrm{max} e^{(t-t_\mathrm{max})/\tau},\;\;\;\;\;\; t<t_\mathrm{max}\\
                  \Delta S_\mathrm{max} e^{(t_\mathrm{max}-t)/1.3\tau},\;\; t>t_\mathrm{max}
                \end{array}
              \right.,
	\label{eq:flares_eq}
$$
where $\Delta S_\mathrm{max}$ is the maximum flux density of the flare in [Jy] calculated on the top of the quiescent level $\Delta S_0$ and observed at the time $t_\mathrm{max}$; $\tau$ is the rise time of the flare. This approach is commonly used for radio data at high frequencies since flux density varies on typically smaller timescales due to opacity. Thus, we performed a flare decomposition for the light curves at 11 and 22 GHz (\autoref{f:RATANall}). To find the number of flares that best describes our data, we applied  the Akaike information criterion \citep[e.g.][]{1974ITAC...19..716A}. It showed that the source light curve is best described by three flares, see the red curve in \autoref{f:neutrino_Fermi_radio_lc} for the 11~GHz data. The fitted parameters of each flare and quiescent flux densities are given in \autoref{tab:flare}.

\begin{table}
	\centering
	\caption{Main parameters of the exponential flares. First row for each frequency indicates the quiescent flux density $\Delta S_0$. The other rows show corresponding parameters of each flare.  }
	\label{tab:flare}
	\begin{tabular}{cccc}
	\hline
	Frequency & Flux density, $\Delta S$ & Peak time,  $t_\mathrm{max}$ & Rise time,  $\tau$ \\
	(GHz)           & (Jy)  & (yr)   & (yr) \\
	\hline
	11 & 0.36 $\pm$ 0.04 & & \\
	   & 1.78 $\pm$ 0.05 & 2020.13 $\pm$ 0.02 & 1.05 $\pm$ 0.05\\
	   & 0.67 $\pm$ 0.05 & 2020.87 $\pm$ 0.01 & 0.23 $\pm$ 0.04\\
	   & 0.37 $\pm$ 0.05 & 2022.01 $\pm$ 0.02 & 0.23 $\pm$ 0.05\\
	\hline
	22 & 0.44 $\pm$ 0.05 & & \\
	   & 1.78 $\pm$ 0.07 & 2019.90 $\pm$ 0.02 & 0.67 $\pm$ 0.05\\
	   & 0.65 $\pm$ 0.05 & 2020.92 $\pm$ 0.01 & 0.19 $\pm$ 0.03\\
	   & 0.37 $\pm$ 0.09 & 2021.99 $\pm$ 0.02 & 0.12 $\pm$ 0.04\\
	\hline
	\end{tabular}
\end{table}

\begin{table*}
    \caption{Estimated all-flavor neutrino flux at $\geq200$~TeV for flares in 2017 and 2021, assuming they are equivalent and both last for a year. Calculations are based on the IceCube detection of one event in 2017, no events in 2021, as well as the Baikal-GVD detection of one event in 2021. Expected number of events for each detector are also provided, together with the probability to observe compatible results given these estimated fluxes. See \autoref{s:discussion} for more details on the motivation and calculations.}
    \label{tab:flarefluxes}
    \centering
    \begin{tabular}{ccc|cccc}
    \hline\hline
  Neutrino &Estimated flux&Consistency &Detector& Effective area& Expected \#events&\\
  spectrum &($10^{-14}$~cm$^{-2}$s$^{-1}$)& probability&& (m$^2$)& (yr$^{-1}$)&\\
    \hline
  \multirow{2}{*}{$E^{-2}$} &\multirow{2}{*}{$5.6^{+7.2}_{-1.8}$}&\multirow{2}{*}{$8\%$} &IceCube& $55.46$& $0.97$& \\
   & &&Baikal-GVD& $3.23$& $0.06$& \\
   \multirow{2}{*}{$E^{-2.5}$}&\multirow{2}{*}{$9.7^{+12.7}_{-3.0}$}&\multirow{2}{*}{$13\%$} &IceCube& $31.27$& $0.96$& \\
   & &&Baikal-GVD& $2.9$& $0.09$& \\
 \hline
\end{tabular}
\end{table*}

\subsection{\textit{Fermi} LAT light curve} 
\label{s:Fermi}

We constructed the gamma-ray photon flux light curve of the source 4FGL~J0509.4+0542, positionally associated with the quasar 0506+056, based on the \textit{Fermi} LAT data taken within the energy range 0.1--300~GeV since 2008 August~4 through 2022 May~11 (\autoref{f:neutrino_Fermi_radio_lc}). We applied the adaptive binning technique \citep{Lott12} with a constant target flux uncertainty of 20\%, and assuming that the source energy spectrum follows a power-law function with photon index $\Gamma=2.079$ \citep{4FGL,4FGLDR2}. The resulting light curve comprises 281 bins with a median bin duration of 17 days, a median Test Statistics \citep{Mattox96} $\mathrm{TS}=64$, and a median photon flux $F_\gamma=6.8\times10^{-8}$~ph~cm$^{-2}$~s$^{-1}$. The source experienced activity periods centered on 2011 and mid-2017, but at the time of the Baikal-GVD cascade neutrino event, the source was in the low-state, with the photon flux
$3.2\times10^{-8}$~ph~cm$^{-2}$~s$^{-1}$, twice as low as the median level.

\section{Discussion}
\label{s:discussion}

We have found that an exceptional Baikal-GVD event, GVD210418CA, having a 99.67\% probability to be of astrophysical origin, is associated with an exceptional source singled out by previous studies. Indeed, not only the energetic, $E=290^{+2010}_{-75}$~TeV \citep[see][Figure~S2]{IceCubeTXSgamma}, neutrino candidate event IC170922A was associated with a gamma-ray flare of TXS~0506+056, but it was the only source claimed by IceCube Collaboration at estimated neutrino energies $E>200$~TeV.
A significant correlation between the neutrino arrival and the gamma-ray activity of a source was found for this case only. It is interesting to estimate the probability that the exceptional Baikal-GVD event coincides with the unique source by chance. All extremely high-energy events with published energies, used to establish TXS~0506+056 as a neutrino source, have $E>200$~TeV, and this threshold was previously chosen for various other analyses \citep[e.g.~by][]{IceCube-1607.08006,neutradio1}. In what follows, we also adopt this energy threshold for the present analysis. Note that lowering the threshold down to $\sim$100~TeV or increasing it up to the energy of the exceptional event, we consider here would not change our numerical estimates considerably. 

\begin{figure*}
    \centering
    \includegraphics[width=0.85\linewidth]{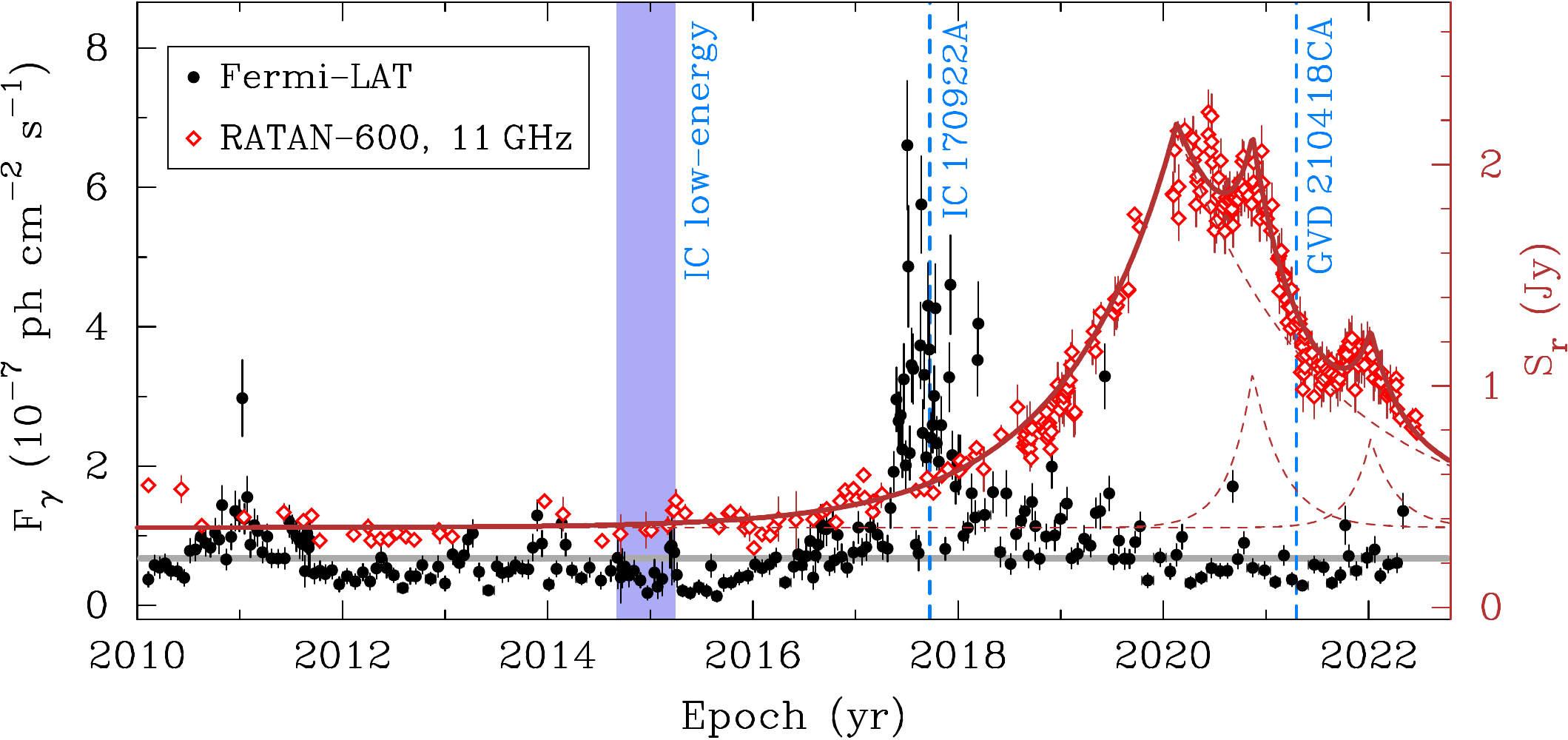}
    \caption{Radio and gamma-ray light curves of TXS~0506+056. Black dots: \textit{Fermi} LAT adaptive binning light curve of the gamma-ray source 4FGLJ0509.4+0542 positionally associated with  TXS~0506+056. Grey horizontal line indicates the median gamma-ray flux. Empty red diamonds: RATAN-600 light curve at 11~GHz. The radio light curve is decomposed by three radio flares depicted by dark-red dashed lines, the sum of which is represented by a thick line.}
    \label{f:neutrino_Fermi_radio_lc}
\end{figure*}

The highest-purity sample of astrophysical neutrino candidates in the Baikal-GVD analysis contains upgoing cascade events only, with zenith angles satisfying $\cos\theta<-0.25$. Only one Baikal-GVD cascade with $E>200$~TeV was detected from this $0.75 \cdot 2\pi$~sr of the sky, the one which we discuss here. The exposure of the detector is not uniform and depends on $\theta$: the Earth is not fully transparent to neutrinos at the highest energies. We use Monte-Carlo simulations of isotropically distributed incoming astrophysical neutrinos of the observed energy, 224~TeV, to take this effect into account. We obtain the chance probability of coincidence, within the 90\,\% confidence level angular-resolution cone of $6.2^\circ$, of the arrival direction of a single event with the position of TXS~0506+056 in the sky, as $p_\mathrm{TXS} = 0.0074$.

The Baikal-GVD detection of the second $E>200$ TeV event from TXS~0506+056, in addition to the IceCube event of 2017, significantly changes the estimate of high-energy neutrino flux from this particular source.
Indeed, while it was possible to obtain an order-of-magnitude estimate of the flux with a single IceCube event, it might be misleading both because of the Eddington bias \citep{EddingtonBias} and due to the lack of knowledge of the typical duration of a flare. Combining the IceCube and Baikal-GVD measurements, the flux can be estimated with an improved uncertainty, and some information about the duration of the flare can be obtained. 

At the time of the Baikal-GVD event from TXS~0506+056 in 2021, eight clusters of GVD were in operation, with the effective area for cascades several times smaller than that of IceCube for the track channel. The observation of one event in GVD and zero events in larger IceCube are reconcilable with the Poisson fluctuations provided the expected number of the observed events is small. This holds only under the assumption of a flaring source, with the duration of flares about a year or shorter. Otherwise, non-observation of events by IceCube would become incompatible with the assumed flux due to its large cumulative exposure.

Assume that the source emits neutrinos in flares with the duration $T \sim 1$~yr, as it is motivated by the radio lightcurve studies \citep{neutradio1,Hovatta2021}. The 2017 IceCube event was reported as an extremely high-energy (EHE) muon-track alert; the corresponding energy-dependent effective area was presented in Figure~S1 of \citet{IceCubeTXSgamma} for the declination bin containing the source of interest. Since 2018, all EHE alerts are included, among others, in the GOLD alert channel \citep{IceCubeGOLD-BRONZE}, and no such event was reported from the direction of TXS~0506+056. The 2021 Baikal event was observed in the cascade mode; the average all-sky effective area for the 7-cluster GVD configuration was published by \citet{Baikal-JETP}. However, the GVD effective areea is a growing function of time in the construction period; here we use the time-dependent exposure estimated for the declination of TXS~0506+056. 

We start from the number of observed events, 1 by IceCube in the 2017 flare, 0 by IceCube, and 1 by Baikal-GVD in the 2021 flare (Baikal-GVD operates since 2018). We find the Poisson maximum likelihood estimate of the $E\geq200$~TeV neutrino flux of such a year-long flare, for two energy spectra of $E^{-2}$ and $E^{-2.5}$. These fluxes are presented in \autoref{tab:flarefluxes}. We demonstrate that the observations are consistent with detector sensitivities by computing the probability to encounter these or less likely detection counts given the assumptions. The probabilities are reasonably high, 8\% and 13\% for the two energy spectra, indicating that the observations are consistent with such flares.

The estimated neutrino flux (\autoref{tab:flarefluxes}) corresponds to ${E^2F(E) \sim 10^{-11}}$~TeV\,cm$^{-2}$\,s$^{-1}$, consistent by the order of magnitude with the IceCube estimates of 2017 and with the gamma-ray flux in the \textit{Fermi} LAT band during the high state of TXS~0506+056 \citet{IceCubeTXSgamma}. However, the \textit{Fermi} LAT flux remains low in the period of detection of the Baikal-GVD event. Together with the earlier observation of the lack of gamma-ray activity during the neutrino flare of the same source in 2014 (\autoref{f:neutrino_Fermi_radio_lc}), this suggests that simple one-zone models are disfavored as an explanation of both gamma-ray and neutrino data. This is in line with strongly constraining results of the population studies, e.g.~\citet{corr-1611.06338Neronov,Murase10percent}, for a recent discussion see \citet{Halzen-no-gamma}.

Contrary, \autoref{f:neutrino_Fermi_radio_lc} reveals an important similarity between the IceCube (2017) and Baikal-GVD (2021) events in terms of the radio observations: both neutrinos arrived at the beginning of prominent radio flares. 
In order to characterize the flaring activity of the blazar in radio waves, we decomposed the two highest-frequency light curves into three flares, see \autoref{a:flares}, \autoref{tab:flare}, and Figures~\ref{f:RATANall},~\ref{f:neutrino_Fermi_radio_lc}.
The neutrino event IceCube-170922A occurred on September 22, 2017, during a slow rising of the strongest radio flare of this source over the past 25 years \citep{2020AdSpR..65..745K} and a large gamma-ray flare (\autoref{f:neutrino_Fermi_radio_lc}). 
The radio flare started in 2015-2016 after a long low-activity state and reached its maximum at 22~GHz at the end of 2019, while lower frequency emission was delayed, as expected from synchrotron opacity. 
The Baikal neutrino event detected on April 18, 2021, may coincide with the beginning of a new radio flare peaking at 2022.0 (see \autoref{f:neutrino_Fermi_radio_lc} and \autoref{tab:flare}). 
We note that it is very complicated to correctly estimate a formal probability of chance coincidence, especially due to its \textit{a posteriori} nature \citep[see also the neutrino--radio--$\gamma$-ray flare coincidence discussion for the blazar J0242+1101 in][]{ANTARES_RFC}. It is expected to be high following results by \citet{neutradio1,Hovatta2021,2022A&A...666A..36L}.

This gives additional support for the statistical association between high-energy neutrinos and compact radio-loud structures in blazars proposed by \citet{neutradio1}, which may give important clues to ultimate understanding of the mechanisms of neutrino production. For instance, both the radio flare and the enhanced rate of neutrino emission may be caused by an increase in the continuous flow of matter falling from the accretion disk onto the supermassive black hole located in the center of the blazar. Part of this matter is thrown away and, after being accelerated to very high energies, forms a relativistic jet. The observed radio emission is caused by the synchrotron radiation of these accelerated particles, and the flare is accompanied by an increase in the density of radiation and matter, providing targets for neutrino production in interactions of accelerated protons.  

VLBI monitoring \citep[e.g.\ the currently ongoing 15~GHz MOJAVE program\footnote{\url{https://www.cv.nrao.edu/MOJAVE}} at the VLBA,][]{lister21} will help to understand what is happening at parsec scales and check if there is a connection between neutrino production, core brightening, a birth, and prorogation of new features through the jet base. See the first results of VLBI monitoring observations after the 2017 high energy IceCube event in \citet{r:ros0506}.

Previous population studies suggested that the number of sources of high-energy neutrinos is large \citep{NeronovEvolution,Finley-2005.02395}. This is true in particular for radio blazars \citep{neutradio2}: statistically significant results were obtained for their large samples only. One does not expect many high-energy neutrino events coming from a single typical source, and this is what was observed. The accumulation of statistics and the contribution from a new experiment, Baikal-GVD, will make these conclusions more refined: multiple high-energy neutrinos start to be recorded from directions to particular blazars, see also \citet{2022ATel15363....1P}. Some of these sources, TXS~0506+056 among them, may be singled out by prolonged periods of higher activity or just demonstrate lucky fluctuations; future higher-exposure studies will be necessary to make the choice.

\section{Summary} \label{s:summary}

The existence of high-energy astrophysical neutrinos has been unambiguously demonstrated, but their sources remain elusive. IceCube, a large neutrino telescope at the South Pole, reported an association of a 290-TeV neutrino with a gamma-ray flare of TXS 0506+056, an active galactic nucleus with a jet pointing to us (a blazar). But population studies suggested that gamma-ray blazars cannot dominate the observed neutrino flux. Contrary, radio blazars, to which class this particular source also belongs, were shown to be associated with IceCube neutrino events with high statistical significance. However, these associations remained unconfirmed with the data of independent experiments, and were based only on track-like events, for which the neutrino energy is poorly reconstructed. Here we report on the detection of a rare neutrino with the estimated energy of 224 TeV from the direction of TXS~0506+056 by the new Baikal-GVD neutrino telescope in April 2021 followed by a radio flare observed by RATAN-600 telescope of the Special Astrophysical Observatory. This event is the highest-energy cascade detected so far by Baikal-GVD from a direction below horizon, where non-astrophysical backgrounds are strongly suppressed. Unlike ice-based IceCube, Baikal-GVD uses liquid fresh water as the target, which allows one to reconstruct neutrino arrival directions with better precision, making it possible to search for high-energy neutrino sources with cascade events. The present result supports previous observations that radio blazars in general, and TXS~0506+056 in particular, are sources of high-energy neutrinos, and opens up the cascade channel for the neutrino astronomy.

\section*{Acknowledgements}

We are grateful to the anonymous referee for constrictive comments and suggestions which have helped to improve the manuscript.
We thank Galina Lipunova and Eduardo Ros for discussions as well as Elena Bazanova for English editing.
This paper is supported by the Ministry of science and higher education of Russia under the contract 075-15-2020-778. 
The work is partially supported by the European Regional Development Fund-Project ``Engineering applications of microworld physics'' (CZ 02.1.01/0.0/0.0/16\_019/0000766) and by the VEGA Grant Agency of the Slovak Republic under Contract No.~1/0607/20.
%

\section*{Data availability}

Data and other material presented in this paper are
available from the corresponding author upon reasonable request.

\medskip\noindent
\textit{Facilities:} Baikal-GVD, \textit{Fermi} LAT, RATAN-600.

\bibliographystyle{mnras}
\bibliography{neutradio}


\bsp	
\label{lastpage}
\end{document}